\documentclass[aps,prl,twocolumn]{revtex4}
\usepackage{graphicx}
\usepackage{amssymb}
\usepackage[dvips]{color}
\newcommand{\bfk}{{\bf k}}
\newcommand{\kp}{{\bf k}'}
\newcommand{\kdp}{{\bf k}''}
\newcommand{\ketk}{|\bfk\rangle}
\newcommand{\ketkp}{|\kp\rangle}
\newcommand{\ketkdp}{|\kdp\rangle}
\newcommand{\bfq}{{\bf q}}
\newcommand{\ha}{\hat a}
\newcommand{\hb}{\hat b}
\newcommand{\hc}{\hat c}
\newcommand{\hS}{\hat S}
\newcommand{\hT}{\hat T}

\newcommand{\keto}{|a\rangle}
\newcommand{\kett}{|b \rangle}
\newcommand{\brao}{\langle a|}
\newcommand{\brat}{\langle b|}
\newcommand{\ketS}{|S\rangle}

\newcommand{\ketT}{|T\rangle}
\newcommand{\braT}{\langle T|}
\newcommand{\braS}{\langle S|}

\begin{document}
\title{On the conversion efficiency of ultracold fermionic atoms to
bosonic molecules via Feshbach resonance sweep experiments}

\author{E. Pazy, A. Vardi, and Y. B. Band}

\affiliation{Department of Chemistry, Ben-Gurion University of the
Negev, P.O.B. 653, Beer-Sheva 84105, Israel}


\begin{abstract}
We explain why the experimental efficiency observed in the conversion
of ultracold Fermi gases of $^{40}$K and $^{6}$Li atoms into diatomic
Bose gases is limited to 0.5 when the Feshbach resonance sweep rate is
sufficiently slow to pass adiabatically through the Landau-Zener
transition but faster than ``the collision rate'' in the gas, and
increases beyond 0.5 when it is slower. The 0.5 efficiency limit is
due to the preparation of a statistical mixture of two spin-states,
required to enable $s$-wave scattering.  By constructing the many-body
state of the system we show that this preparation yields a mixture of
even and odd parity pair-states, where only even parity can produce
molecules.  The odd parity spin-symmetric states must decorrelate
before the constituent atoms can further Feshbach scatter thereby
increasing the conversion efficiency; ``the collision rate'' is the
pair decorrelation rate.
\end{abstract}


\maketitle

The formation of quantum-degenerate molecular gases by association of
quantum-degenerate atomic gases has been the subject of intense
theoretical deliberation in recent years \cite{Julienne98,Drummond98,%
Timmermans99,Javanainen99,Mies00,Vardi01,Holland01}.  Experimental
efforts soon followed, demonstrating the association of an atomic
Bose-Einstein condensate (BEC) into ground-state molecules either via
a stimulated Raman transition \cite{Wynar00} or through a
magnetic-field controlled Feshbach resonance \cite{Donley02}.

In four recent experiments a Fermi gas of atoms was converted into an
ultracold Bose gas of molecules by adiabatic passage through a Feshbach
resonance \cite{Regal03,Strecker03,Cubizolles,Greiner}.  In two of
these experiments \cite{Regal03,Greiner} ultracold $^{40}$K$_{2}$
molecules were produced from a quantum degenerate Fermi gas of $^{40}$K
atoms, whereas in Ref.~\cite{Strecker03,Cubizolles} $^{6}$Li atoms were
converted to diatomic molecules.  An interesting feature of these
studies is that when the sweep rate through the Feshbach resonance was
comparable to the background elastic scattering rate, yet slow with
respect to the atom-molecule coupling rate, a maximum atom-molecule
transfer efficiency of 50\% was reported \cite{Regal03,Strecker03}.
However, when the sweep rate was much slower, approaching ``close to
thermal equilibrium'' conditions, higher conversion efficiency was
attained \cite{Cubizolles,Greiner}.

\begin{figure}
\centering
\includegraphics[scale=0.40,angle=0]{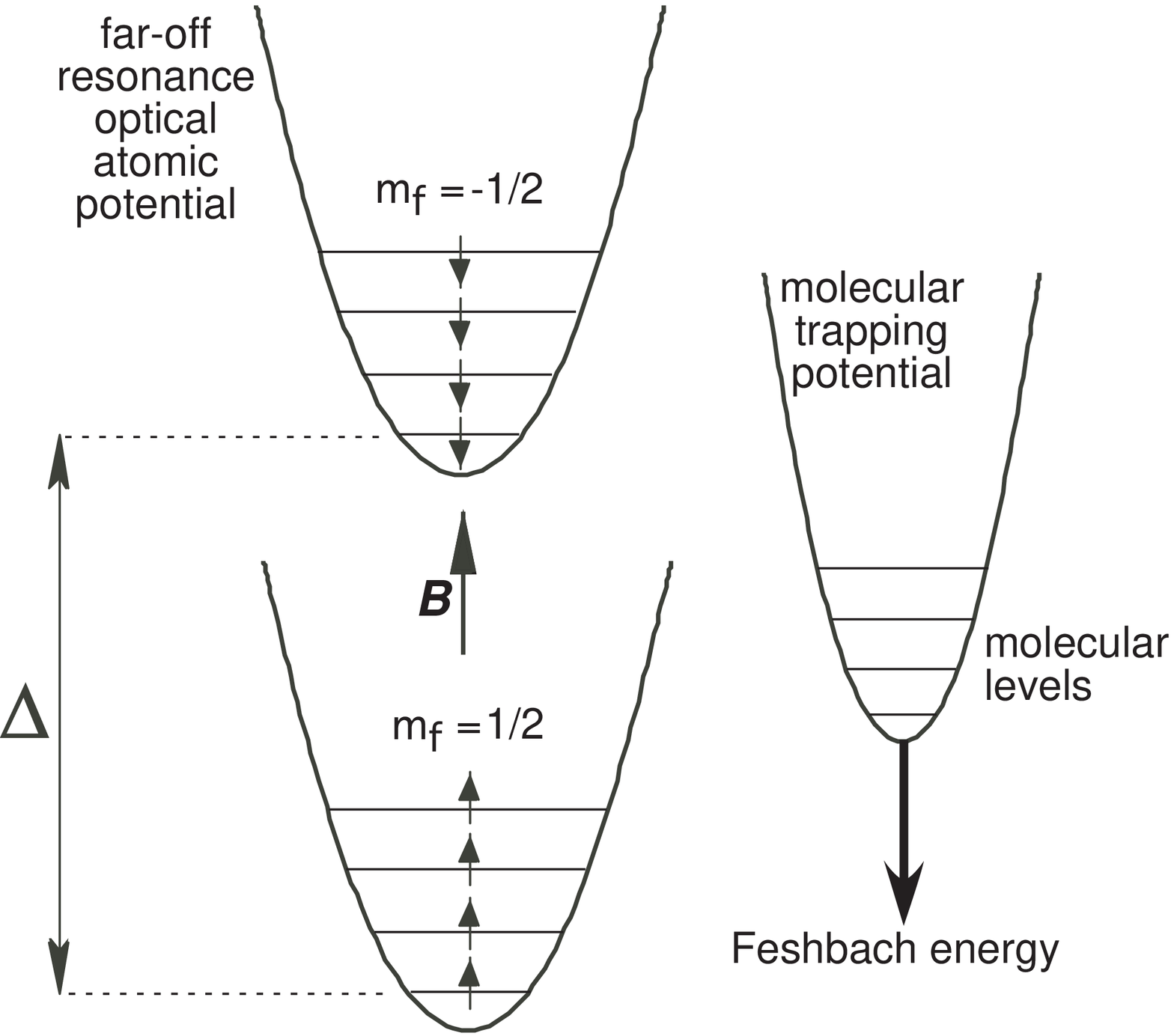}
\caption{Schematic drawing of the trapping potentials for atoms with
different spin projections, $m_a$ and $m_b$ (explicitly labeled
$\uparrow$ and $\downarrow$ as appropriate for the $^{6}$Li case), and
for diatomic molecules (on the right side of the figure) whose
potential is swept downward in energy as a function of time.  $\Delta$
is the Zeeman energy splitting.  The molecular potential force
constant is twice as large as that of the atoms, but the energy level
spacings of the atoms and molecules are equal.}
\label{fig1}
\end{figure}

In this Letter we show that the observed 50\% saturation of the
atom-molecule conversion efficiency in the fast (yet adiabatic)
Feshbach sweep regime is a result of the initial state preparation.
The fermionic atoms in the experiments of
Refs.~\cite{Strecker03,Regal03,Cubizolles, Greiner} are prepared in
statistical mixtures of two spin-states in order to enable their
$s$-wave scattering.  Elastic-scattering collisions between atoms in
these different spin-states results in cooling of the trapped gas as
hot atoms leave the trap.  This procedure effectively splits the
many-body system into two subsystems corresponding to two different
spin-states, as shown in Fig.~\ref{fig1}.  If we consider pairs of
atoms, one from each spin-state (where spin means total-spin --
$|\tilde{f} \, m_{f} \rangle$ -- not electronic spin \cite{f_note}),
half the pairs are antisymmetric spin-states (for $^{6}$Li these are
spin-singlets) and half are symmetric spin-states (for $^{6}$Li,
spin-triplets).  More explicitly, the reduced two-particle density
matrix obtained by tracing out all but one particle of one spin-state
and another particle of the other spin-state contains 50\%
spin-antisymmetric (spin-singlet for $^{6}$Li), interacting via an
$s$-wave (even parity spatial state), and 50\% spin-symmetric (e.g.,
the triplet spin-symmetric superposition of the form $|\uparrow
\downarrow \rangle + |\downarrow \uparrow \rangle$ for $^{6}$Li),
interacting via a $p$-wave (odd parity spatial state).  Sweeping
through the resonance sufficiently slowly so the Landau-Zener
transition \cite{LZ} is traversed adiabatically \cite{Mies00},
spin-singlet pairs are converted to molecules.  However, since the
Feshbach is an $s$-wave resonance, spin-triplet pairs cannot Feshbach
scatter.  Once a triplet state is formed, its constituent atoms can
only undergo triplet/p-wave collisions with other atoms until they are
decorrelated.  Hence, if collisional dephasing is slow with respect to
the sweep rate, colliding pairs of atoms form either spin-singlets
(eventually producing a molecule via the Feshbach resonance sweep) or
spin-triplets (producing a correlated pair that cannot Feshbach
scatter).  If the sweep rate is slower than the time required for
spin-triplet states to decorrelate, the conversion efficiency can grow
beyond 50\%.

Both the $^{6}$Li and $^{40}$K Feshbach experiments involve two atomic
spin-states $|\chi\rangle \equiv |\tilde{f}_{\chi} m_{\chi} \rangle$ 
with
$\chi=\{a,b\}$.  An external magnetic field generates a large Zeeman
splitting $\Delta$ between the spin-states $\keto$, $\kett$, as
depicted in Fig.~\ref{fig1}.  We consider the Feshbach atom-molecule
coupling Hamiltonian
\begin{equation}
\label{hf}
\hat H_F=g \sum_{\bfq,\bfk} \ha^\dag_{\bfq/2-\bfk}
\hb^\dag_{\bfq/2+\bfk} \hc_\bfq + {\mathrm{h.c.}} \,,
\end{equation}
where $g$ is the coupling coefficient, $\ha^\dag_{\bfq/2-\bfk}$ and
$\hb^\dag_{\bfq/2+\bfk}$ are the usual atomic creation operators for
spin-states $\keto$ and $\kett$ respectively, and $\hc_\bfq$ is
the molecular annihilation operator.  Written explicitly for fermionic
operators $\ha_\bfk$, $\hb_\bfk$, and bosonic $\hc_\bfq$,
the Hamiltonian (\ref{hf}) takes the form:
\begin{equation}
\label{hff}
\hat H_F= {g \over 2} \sum_{\bfq,\bfk}
\left(\ha^\dag_{\bfq/2-\bfk}\hb^\dag_{\bfq/2+\bfk}
-\hb^\dag_{\bfq/2-\bfk} \ha^\dag_{\bfq/2 + \bfk}\right) \hc_\bfq
+{\mathrm{h.c.}} \,,
\end{equation}
having used the anti-commutation relation
$\{\ha_\bfk^\dag,\hb_{\bfk'}\}=0$.

It is evident that the fermionic Feshbach coupling Hamiltonian
(\ref{hff}) is antisymmetric under spin permutation and that it has
even parity.  It only allows for the association of atom pairs
colliding with even parity partial waves (effectively restricted at
ultracold temperatures, to $s$-wave scattering) with an antisymmetric
spin-state (a spin-singlet in the $^{6}$Li case).  Defining the
commuting operators
\begin{eqnarray}
\hS_{\bfq} &=& \sum_\bfk\left(\ha_{\bfq/2-\bfk}
\hb_{\bfq/2+\bfk}-\hb_{\bfq/2-\bfk}\ha_{\bfq/2+\bfk}\right)\,, \\
\hT_{\bfq} &=& \sum_\bfk\left(\ha_{\bfq/2-\bfk}\hb_{\bfq/2+\bfk}+
\hb_{\bfq/2-\bfk}\ha_{\bfq/2+\bfk}\right)\,,
\end{eqnarray}
the Feshbach coupling Hamiltonian for fermionic atoms simply reads
\begin{equation}
\hat H_F = {g \over 2} \sum_\bfq \hS^\dag_\bfq \hc_\bfq +
{\mathrm{h.c.}} \,.
\label{hfs}
\end{equation}
We note parenthetically that for bosonic operators $\ha_\bfk$,
$\hb_\bfk$, Eq.~(\ref{hf}) assumes the form $\hat H_F = {g
\over 2}\sum_\bfq \hT^\dag_\bfq \hc_\bfq+{\mathrm{h.c.}}$.

As discussed in Refs.~\cite{Strecker03,Regal03}, if all the fermions
are in a pure superposition of internal spin-states, e.g., $\alpha
\keto + \beta \kett$, the colliding atoms will have a symmetric
spin-state and therefore any pair is only allowed to interact through
odd-parity partial waves (e.g., $p$-wave).  In order to enable the
$s$-wave scattering of atoms, a statistical mixture of two Fermi gases
corresponding to two spin-states is obtained by rapidly driving the
Zeeman transition in the presence of a magnetic field gradient.  As
different atoms are driven at different frequencies, the above process
results in an inhomogeneous broadening of the transition.  Since all
coherence between the two spin-states is thus destroyed, the initial
state of the system, after cooling, is best described by a density
matrix composed of an {\em incoherent mixture} of two decoupled Fermi
gases, one for each spin-state:
\begin{eqnarray}
\rho_0^N &=& {\mathcal A} \left[\left(\rho_{N/2}(a,T) \, \keto^{(N/2)}
\,
^{(N/2)}\brao
\right) \otimes \right. \nonumber\\
&~& \left. \left( \rho_{N/2}(b,T) \, \kett^{(N/2)} \, ^{(N/2)}\brat
\right) \right] {\mathcal A}^{-1} \,,
\end{eqnarray}
where $\rho_{N/2}(\chi,T)$ denotes the motional part of the density
matrix for $N/2$ atoms with spin projection $\chi=\{a,b\}$ at
temperature $T$, and ${\mathcal A}$ is the anti-symmetrization operator
of all the particles.

\begin{figure}
\centering
\includegraphics[scale=0.4]{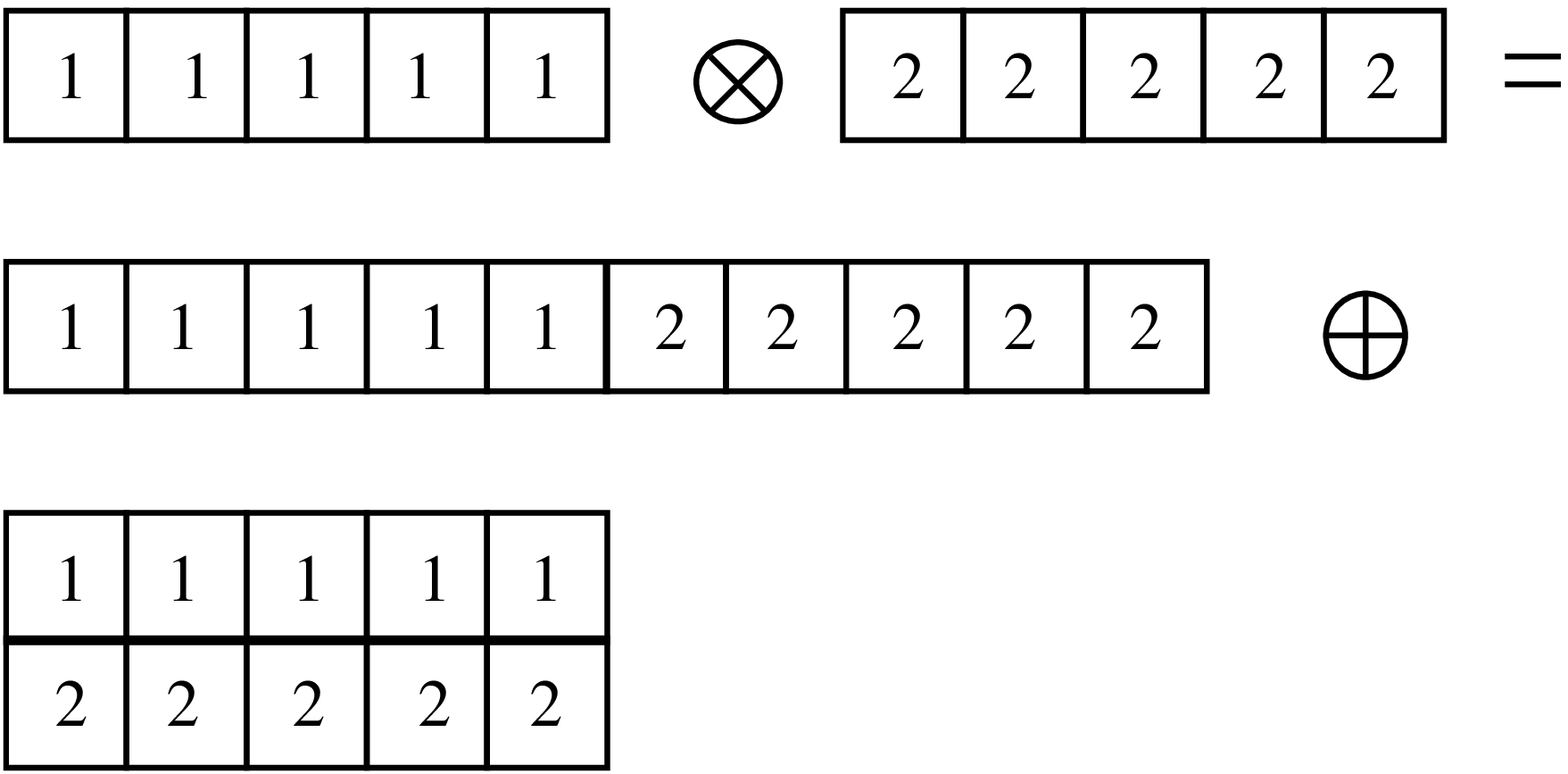}
\caption{Spin degrees of freedom Young Tableaux corresponding to a
totally spin-symmetric configuration for each of the Fermi seas
(each row tableau corresponds to a different spin projection $m_a$
denoted by 1 and $m_b$ denoted by 2) and its decomposition into a
totally symmetric $F=N/2$ and a $F=0$ many-body configuration.  Young
tableaux for spatial degrees of freedom are obtained by interchanging
rows and columns.}
\label{fig:tab}
\end{figure}

Before the Feshbach sweep begins, cooling of the two Fermi gases
induces pair-wise and many-body correlations.  As opposed to a thermal
gas, the probability for forming a molecule in a single Feshbach
collision is determined by the symmetry of the many-body state.  It is
instructive to explicitly construct the many-body state of the cooled
system.  To do so we start by anti-symmetrizing each of the wave
functions for the two Fermi seas separately.  Since atoms in each
Fermi sea have the same spin projection, their many-body wave function
is a tensor product of an odd parity Slater determinant for the
spatial degrees of freedom and a totally symmetric spin configuration.
It is evident from the construction of the pertinent Young tableaux
(see Fig.~\ref{fig:tab}) that the anti-symmetrization of the {\em two}
Fermi seas can be done in only two ways: a totally antisymmetric
spatial wave function times a spin-symmetric configuration $|F=N/2,
M_F=0 \rangle$ (the row Young tableau on the right hand side of the
equation in Fig.~\ref{fig:tab}), and a state which is
spin-antisymmetric upon exchange of particles of different spin
corresponding to $|F=0, M_F=0 \rangle$ (the second tableau on the
right hand side of the equation in Fig.~\ref{fig:tab}).  This is the
many-body analogy of the two spin 1/2 case in which the two body state
naturally falls into two classes: symmetric or antisymmetric in the
spatial coordinates corresponding to spin antisymmetric (singlet) and
spin symmetric (triplet) states respectively, such that the total
fermionic wave function is antisymmetric with respect to exchange of
the two particles.

Let us consider the reduced two-particle density matrix $\rho_0^{(2)}
= \mathrm{Tr}_{(N-2)} \left(\rho_0^N \right)$ obtained by tracing out
all but one particle with spin projection $a$ and another particle
with spin projection $b$.  It is readily seen from the Young tableaux
in Fig.~\ref{fig:tab} that tracing out all but two of the particles,
one with spin up and the other with spin down, will leave a two
particle reduced density matrix which is an equally weighted sum of a
spin symmetric (two box row tableau) and a spin antisymmetric (two box
column tableau) atomic pair.  Therefore, if a given atom pair is
spatially antisymmetric, then each of the atoms comprising this pair
will be spatially antisymmetric with any other atom of opposite spin
due to the symmetry of the many-body state.  This antisymmetrization
induces atomic correlations and therefore determines collision
probabilities.  The spin part of $\rho_0^{(2)}$ is identical for all
atomic pair-states, and is obtained by projection on to the momentum
states of any given colliding pair,
\begin{eqnarray}
&& \langle \bfq/2-\bfk|\langle\bfq/2+\bfk|\rho_0^{(2)}|\bfq/2+\bfk
\rangle|\bfq/2-\bfk\rangle = \nonumber \\
&& = \frac{1}{2}\left(\keto\kett\brat\brao + \kett\keto\brao\brat\right)
    \nonumber \\
&& = \langle \bfq | \, \frac{1}{2}\left(\ketS\braS+\ketT\braT\right) |
\bfq \rangle \,.
\label{eq:densitymat}
\end{eqnarray}
where $|\bfq/2 \pm \bfk \rangle$ is the single-particle spatial wave
function with momentum $\bfq/2 \pm \bfk$, $\ketT = \left(\keto\kett +
\kett\keto\right)/\sqrt{2}$ is a spin symmetric triplet state, and
$\ketS = \left(\keto\kett - \kett\keto\right)/ \sqrt{2}$ is a spin
antisymmetric (singlet) state.  In the middle line of
Eq.~(\ref{eq:densitymat}) we have omitted the projection onto momentum
states to simplify the notation.  Equation (\ref{eq:densitymat})
demonstrates the simple result that in the absence of coherence
between colliding particles in different spin-states, collisions with
a spin-triplet (via a $p$-wave) are just as probable as collisions
with a spin-singlet (via an $s$-wave).  In the above expressions, an
explicit orbital-spin form of the triplet state is given by
\begin{eqnarray}
|T\rangle &=& [|\phi_a(1) \phi_b(2) \rangle - |\phi_b(1) \phi_a(2)
\rangle] \times \nonumber \\
&& [|f_{a} m_{a}, f_{b} m_{b} \rangle + |f_{b} m_{b}, f_{a}
m_{a} \rangle] \,,
\end{eqnarray}
where $\phi_a(i)$ is the orbital for the $i$th atom of species $a$.
For example, for the $^6$Li case,
\begin{equation}
|T\rangle = [|\phi_a(1) \phi_b(2) \rangle - |\phi_b(1) \phi_a(2)
\rangle] \, \, [|\uparrow \downarrow \rangle + |\downarrow \uparrow
\rangle] \,.
\end{equation}
Similarly for $|S\rangle$, but with the signs reversed.

We can describe the process of adiabatically scanning the magnetic
field from high to low field in terms of a Landau-Zener transition
between a pair of free atoms, one from each Fermi gas, into the
resonant vibrational level of the diatomic molecule.  Since only the
spin antisymmetric, even parity, pairs are coupled to molecules by the
Feshbach Hamiltonian (\ref{hfs}), the maximal anticipated conversion
efficiency is 0.5.  The spin-symmetric pairs cannot interact via the
Feshbach resonance and therefore cannot be converted to molecules.

As long as pair correlations are maintained, constituent atoms of
the spin symmetric pairs remaining after Feshbach coupling cannot
interact with another atom to produce a spin-antisymmetric state via
elastic scattering collisions.  This can be demonstrated by
considering the state formed by the tensor product of a spin-symmetric
pair with another fermion and anti-symmetrizing:
\begin{eqnarray}
{\cal A} \{ \left[\ketk_1\ketkp_2-\ketkp_1\ketk_2\right]
\left(\keto_1\kett_2+\kett_1\keto_2\right)\otimes\ketkdp_3\keto_3 \}
\nonumber \\
=\left[\ketk_1\ketkp_2-\ketkp_1\ketk_2\right]
\left(\keto_1\kett_2+\kett_1\keto_2\right)\otimes\ketkdp_3\keto_3
\nonumber \\
-\left[\ketk_3\ketkp_2-\ketkp_3\ketk_2\right]
\left(\keto_3\kett_2+\kett_3\keto_2\right)\otimes\ketkdp_1\keto_1
\nonumber \\
-\left[\ketk_1\ketkp_3-\ketkp_1\ketk_3\right]
\left(\keto_1\kett_3+\kett_1\keto_3\right)\otimes\ketkdp_2\keto_2 \,.
\label{eq:antisymm}
\end{eqnarray}
Tracing out one of the particles results in only spin symmetric pair
states, implying that no spin antisymmetric collisions are allowed
until the spin symmetric pair decoheres.  Furthermore, using similar
argumentation, it is also evident that pairs of spin symmetric pairs
can not interact in such a way that two of the four atoms scatter as a
spin antisymmetric, even parity, pair.  Eq.~(\ref{eq:antisymm}), is
just a specific three particle example of the general case which is
apparent by considering the Young tableaux in Fig.~\ref{fig:tab} from
which one can infer that all two particle states are either odd parity
spin-symmetric for all atoms, or even parity spin-antisymmetric for
all atoms with different spin projections.  It explicitly depicts the
case when one has to combine a two box row tableau with a further box
tableau, and since the states with the same spin projection need to be
symmetric this can only be done in one way.  Therefore, until the spin
symmetric states decohere, allowing a finite probability for spin
antisymmetric (even parity) collisions, no further molecules can be
made from their constituent atoms.

Slowing the sweep rate through the Feshbach resonance, $\gamma_f =
\dot{B} \, (\Delta B)^{-1}$ where $B(t)$, is the time-dependent
magnetic field and $\Delta B$ is the Feshbach resonance width, to the
point where it is smaller than the pair decorrelation rate
$\gamma_{st}$ for triplet to singlet transitions, allows the two
particle spin symmetric pairs not swept by the Feshbach resonance to
be destroyed by collisions.  Consequently, further molecule-producing
$s$-wave/spin-antisymmetric collisions can take place and higher than
50\% efficiency may be obtained as reported by Cubizolles {\it
et.~al.} \cite{Cubizolles} and Greiner {\it et.~al.} \cite{Greiner}.

\begin{figure}
\centering
\includegraphics[scale=0.4,angle=90]{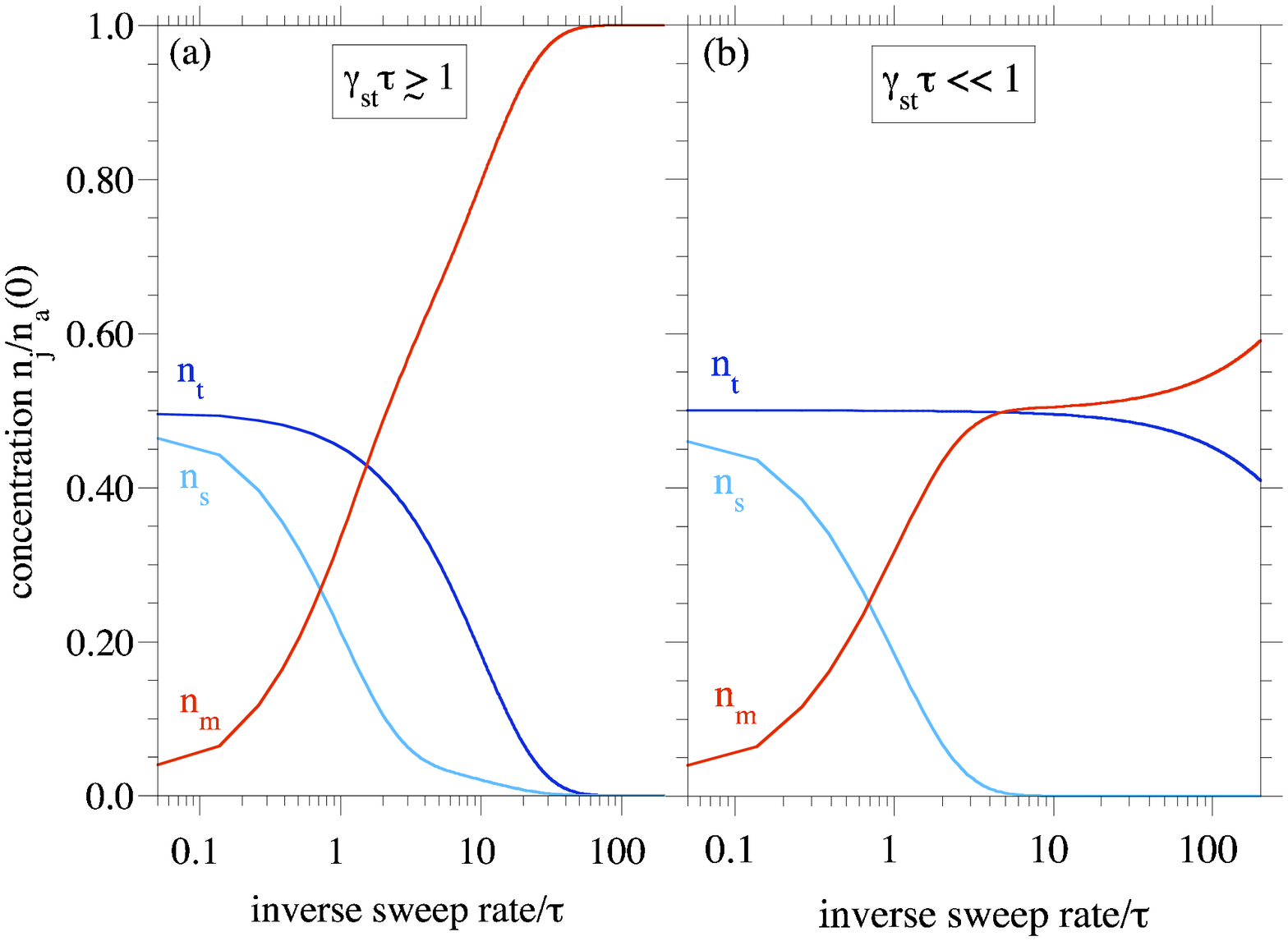}
\caption{Concentrations of molecules $n_m$, singlet-like atomic pairs
$n_s$, and triplet-like atomic pairs $n_t$ versus inverse sweep rate
for two limiting values of the pair decorrelation rate: (a)
$\gamma_{st}\tau \ge 1$, (b)$\gamma_{st}\tau \ll 1$.}
\label{fig3}
\end{figure}

In Fig.~\ref{fig3} we schematically illustrate the expected final
populations of singlet and triplet pairs, and molecules as a function
of the Feshbach sweep rate.  When the sweep rate $\gamma_f$ is fast
compared to the elastic scattering rate in the gas $\tau^{-1}$, the
conversion efficiency is limited to less than 50\% due to the
Landau-Zener nature of the process \cite{LZ,Mies00} and the fact that
only the spin-singlet pairs can participate.  When the sweep rate is
slow compared to $\tau^{-1}$, the conversion efficiency depends on
whether the time spent sweeping across the resonance is fast or slow
compared to the pair decorrelation time $\gamma_{st}$.  If
$\gamma_f\gg\gamma_{st}$ only the spatially-symmetric singlet portion
is converted, leaving the spatially antisymmetric triplets
unperturbed, and one is limited to 50\% conversion.  If $\gamma_f$ is
slow compared to $\gamma_{st}$, then as one sweeps, triplet and
singlet states interconvert and one has the possibility of converting
all the atoms.  Consequently, we distinguish between two cases.  When
$\gamma_{st}\tau\ge 1$, i.e., the decorrelation process is fast
compared to the Feshbach sweep rate, the conversion efficiency ranges
from zero to unity, increasing with decreasing sweep rate as shown in
Fig.~3a.  The adiabaticity criterion for achieving unit conversion
efficiency is in this case $\gamma_f\tau \gg 1$.  On the other hand,
when the decoherence rate is slow compared to the sweep rate,
$\gamma_{st}\tau \ll 1$, the new timescale $\gamma_{st}$ plays a
cruicial role in determining conversion efficiency, as depicted in
Fig.~3b.  For fast sweep rates, $\gamma_f\tau\gg 1$, adiabaticity is
violated and the conversion efficiency is low.  For slower sweeps such
that $\gamma_{st}\tau \ll \gamma_f\tau \ll 1$, two-body adiabaticity
is attained but since the sweep is still too fast to allow
triplet-singlet conversion the conversion efficiency stagnates at a
broad 50\% plateau as in the experiments of
Refs.~\cite{Regal03,Strecker03}.  Only when the sweep rate is the
slowest timescale in the system, $\gamma_f\tau \ll \gamma_{st}\tau \ll
1$, can the conversion efficiency limit to 100\%, as expected from a
truly many-body adiabatic sweep (over 50\% observed in
\cite{Cubizolles,Greiner}).

In summary, we have explained the 0.5 limited atom to molecule transfer
efficiency observed in experiments for fermionic atoms involving an
adiabatic sweep through a Feshbach resonance.  By constructing the
many-body state we demonstrate that half the atoms will be in a
spatially antisymmetric state and therefore will not collide, whereas
the other half will be in a spin symmetric state in which their
collision can lead to the formation of a molecule.  If the sweep is
slow enough, half the atoms will convert to molecules, and the
remaining atoms will not interact since any pair will be in a
spatially antisymmetric state.  We note that even though all remaining
pairs are in a triplet like correlated state, spontaneous transverse
magnetization will not develop since the total $m_{f}$ of any pair is
zero, and this is a good quantum number.  Significant modification of
the collective excitation spectrum \cite{Vichi} and the free expansion
profile of the remaining atomic Fermi gas are anticipated due to the
complete lack of interactions, particularly in the vicinity of the
Feshbach resonance.  We further note that the remaining noninteracting
Fermi gas of atoms can be made to interact by repeating the initial
mixing procedure.  Measurement of the conversion efficiency as a
function of sweep rate in the whole range from very fast to very slow
should serve to conclusively verify or contradict the picture of the
conversion efficiency painted here.

\begin{acknowledgments}
We gratefully acknowledge stimulating discussions with P. S. Julienne,
Y. Avishai, J. Anglin, R. G. Hulet and M. Trippenbach.  This work was
supported in part by grants from the U.S.-Israel Binational Science
Foundation (grant Nos.~2002214, 2002147), the Minerva Foundation
through a grant for a Minerva Junior Research Group, and the Israel
Science Foundation for a Center of Excellence (grant No.~8006/03).
\end{acknowledgments}


\begin{thebibliography}{99}
\bibitem{f_note}
The atomic state in a strong Zeeman field is $|\tilde{f} \, m_{f}
\rangle = |\tilde{f} \, m_{f} (f,j,i,m_j,m_i)\rangle$ where $m_{f}$ is a
good quantum number, but $f$ is not; $\tilde{f}$ is a quantum number
that specifies a particular diagonalized state in the Paschen-Back
limit.
\end{thebibliography}
\end{document}